\documentclass[aps,pre,preprint,groupedaddress,showpacs]{revtex4}
\newif\ifpdf            
\ifx\pdfoutput\undefined \pdffalse \else \pdftrue \fi
\ifpdf 
\usepackage[pdftex]{graphicx}
\pdfcompresslevel9 \DeclareGraphicsExtensions{.jpg,.pdf,.png,.mps}
\else 
\usepackage[dvips]{graphicx}
\DeclareGraphicsExtensions{.eps,.ps} \fi
\usepackage{bm}
\usepackage{color}

\begin{document}

\title{Mechanically induced helix-coil transition in biopolymer networks}

\author{S.~Courty, J.L.~Gornall and E.M.~Terentjev}

\affiliation{Cavendish Laboratory, University of Cambridge,
Madingley Road, Cambridge, CB3 0HE, U.K. }

\date{\today}

\begin{abstract}

The quasi-equilibrium evolution of the helical fraction occurring
in a biopolymer network (gelatin gel) under an applied stress has
been investigated by observing modulation in its optical activity.
Its variation with the imposed chain extension is distinctly
non-monotonic and corresponds to the transition of initially
coiled strands to induced left-handed helices. The experimental
results are in qualitative agreement with theoretical predictions
of helices induced on chain extension. This new effect of
mechanically stimulated helix-coil transition has been studied
further as a function of the elastic properties of the polymer
network: crosslink density and network aging.

\end{abstract}


\maketitle

\section{Introduction}

Biopolymer networks have been widely studied over the recent years
because of their many practical applications in fields such as
food, biomedical and even photographic industries. For example,
they have been extensively used for capsules, adsorbent padding,
tissue regeneration and also as neutral density filters for
optics. More fundamentally, biological networks
(e.g.~cytoskeletons with actin and microtubules) play an important
role in cells by providing a structural framework which is
responsible for the mechanical stability and the locomotion of the
whole cell, as well as intracellular transport processes
\cite{cell}. Collagen networks form the core structure of bone,
ligaments and other living tissues. They have suscitated a high
interest for a growing community of researchers at the interface
of physics and biology \cite{MacKintosh1995,Weitz1999}.

Biopolymer networks also offer new alternatives to explore the
relationship between the structure and responses at molecular
length-scales. Certain homopolypeptides form regular $\alpha$-helices
under appropriate conditions. In this case, the molecular
configurations are well understood according to the Zimm-Bragg model
\cite{Zimm1959} (and its many subsequent modifications
\cite{Grosberg1994}). This model assumes that each segment along the
polymer chain has access to only two conformational states, a
random-coil state and a helical state, where the particular residue
forms a hydrogen bond with specific other residues at a certain
distance along the backbone. By modifying the end-to-end distance $R$
of individual chains, the equilibrium state between the helical and
the random-coil segments can evolve and a coil to helix transition can
be induced. To determine the relationship between the helical state
and the chain end-to-end distance, one needs to control the distance
$R$ and simultaneously measure the helical content. With the recent
development of single-molecule force spectroscopy (SMFS), it has been
possible to measure the forces generated by biopolymers and their
response to applied extension forces. These experiments can tackle the
question of how chiral biopolymers are held in their native state, and
their pathways of folding and unfolding.  Many macromolecules have
been studied using SMFS including biopolymers such as DNA, xanthan and
dextran \cite{Perkins1994,Hongbin1998,Rief1997} and synthetic polymers
such as poly(ethylene-glycol) \cite{Oesterhelt1999}.  However, while
SMFS has revealed the spectacular force-extension curves, information
about the structural transitions which occur on extension of these
chains (e.g.~the degree of helicity in a single molecule) is difficult
to obtain.  Theoretical modeling of the behavior of single biopolymer
molecules on extension has been hampered by this lack of access to
direct information on structural changes and has only been done in
specific cases \cite{Rief1998}.

In this paper, we investigate the internal structure of a
biopolymer network of helix-forming chains, and its response to an
applied stress, using a new macroscopic approach combining
mechanical and optical methods. We find a spectacular
non-monotonic relationship between the helical content and an
externally imposed deformation, which translates to the change in
end-to-end distance of network chains. In particular, we identify
the effect of induced helix formation at a critical strain and
further study this new effect as a function of the elastic
properties of the network, using gelatin (denatured collagen) as a
model system.

\section{Theoretical background} \label{sec:theo}

According to the Zimm-Bragg model, each segment along a polymer chain
only has access to two states: the random coil-like unbound state, and
the helical state \cite{Zimm1959}. The state of the polymer can
therefore be described by the sequence of these units along the chain,
[$hhcchcccch\ldots$], where $h$ stands for helix and $c$ for coil. If
the polymer, or any part of it, is in the helical state, then its
effective length is shortened by a factor $\gamma$, which is about 0.4
for a typical polypeptide. Since the hydrogen bonds prevent free
rotation, the persistence length of the helical state is much
increased compared to that of the coil, $\sim 200$~nm and $\sim
1.8$~nm respectively \cite{Grosberg1994,Poland1970}. The number of
available configurations also decreases on helix formation, and
therefore the gain $\Delta{h}$ in potential energy per monomer by
forming hydrogen bonds competes with the associated loss in entropy
$\Delta{s}$. The balance between these two forces can be expressed in
terms of the free energy per unit monomer, $\Delta{f} = \Delta{h} -
T\Delta{s}$, where $T$ is the temperature. Monomers located at the
ends of helical domains suffer a reduction in entropy but do not form
hydrogen bonds. Therefore, monomers on the boundary between a helical
and coil domain have an increased free energy of $\Delta{f_{t}} =
-\Delta{h}$ compared to a monomer in the helical state. The effect of
this term is to suppress domain boundaries; the helix-coil transition
in this model is cooperative.  The parameters $\Delta{f}$ and
$\Delta{f_{t}}$ are usually expressed in terms of the Zimm-Bragg
parameters $s$ and $\sigma$ \cite{Zimm1959},
\begin{equation}
s = \textrm{exp}(-\beta{\Delta{f}}) ; \ \ \  \sigma =
\textrm{exp}(-2\beta{\Delta{f_{t}}}) \label{eq:sigma}
\end{equation}
where $\beta=1/(k_{B}T)$.  The factor $2$ in (\ref{eq:sigma})
takes into account the fact that a helical domain has two
boundaries with coil regions. Having identified the microscopic
states of a chain the average helical fraction of an ensemble of
polymers $\langle{\chi}{\rangle}$ can be determined by the usual
statistical mechanics calculation maximizing the partition
function,
\begin{equation}
Z = {\sum_{s_{1}s_{2}\ldots{s_{N}}}^{}}
e^{-\beta{F}{([s_{1}s_{2}\ldots{s_{N}}])}}
\end{equation}
where $s_{i}$ is either $c$ or $h$ and the sum is performed over
all possible states of the variables $s_i$, $i=1\ldots{N}$. The
total free energy $F$, which depends on [$s_i$], is the sum of the
monomer free energies given by $\Delta{f}$ and $\Delta{f_{t}}$.
This procedure gives a higher statistical weight to configurations
with a large number of small domains rather than a small number of
large domains; for a fixed number of helical domains there are
more possible rearrangements that can occur without a change in
energy (provided helix still follows coil and vice versa) for a
large number of small domains.

\begin{figure}
\resizebox{0.5\textwidth}{!}{\includegraphics{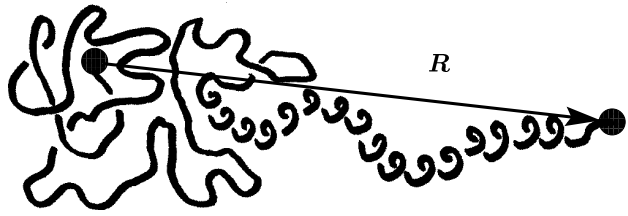}}
 \caption{The sketch of the chain with helical and coil fractions,
at constrained end-to-end distance $R$.} \label{diblock}
\end{figure}

Recently, Tamashiro and Pincus \cite{Pincus2001}, and Buhot and
Halperin \cite{Buhot2002} have shown that by imposing an extension
on such a helix-forming chain, which is kept above but close to
the spontaneous helix-coil transition, one can stimulate the
helical state (which corresponds to a natural enthalpy well
$\Delta h$) by reducing the randomizing effect of chain entropy
($\Delta s <0$) due to the stretching of its ends. The problem can
be simplified by neglecting the mixing entropy without changing
the outcome qualitatively, and only slightly quantitatively, i.e.
assuming the helical and coil domains are separated into two
blocks with a fixed interface energy $\Delta f_t$,
Fig.~\ref{diblock}. The free energy of the chain in this case is
then given by
\begin{equation}
F_{ch} = \chi{N}\Delta{f} + 2\Delta{f_t} +
\frac{3(R-\gamma{aN\chi})^2}{2(1-\chi)Na^2}\label{free energy}
\end{equation}
where $\Delta{f}=\Delta{h}-T\Delta{s} >0$ and $N$ is the total
number of monomers of length $a$. The first two terms give the
free energy of a helix, with its interface, and they are positive
since under our assumption the chain is above its spontaneous
coil-helix transition. The last term represents the Gaussian
entropic free energy describing the coil fraction with an
end-to-end distance $R$ minus the distance bridged by the helical
part of the chain, with the reduced contour length available to
the coil. Here $\chi \in (0,1)$ is the helical content and
$\gamma$ the geometric factor accounting for the helix length, per
monomer. Depending on parameters, this model predicts that the
helical content of the chain, $\chi (R)$, may increase on
stretching (with or without a threshold), finding an optimal
balance between the two competing factors increasing the total
$F_{ch}$. Eventually all the chain is in the $\alpha$-helix
($\chi=1$) and on further extension one would of course force the
helix to unwind.
\begin{figure}
\resizebox{0.6\textwidth}{!}{\includegraphics{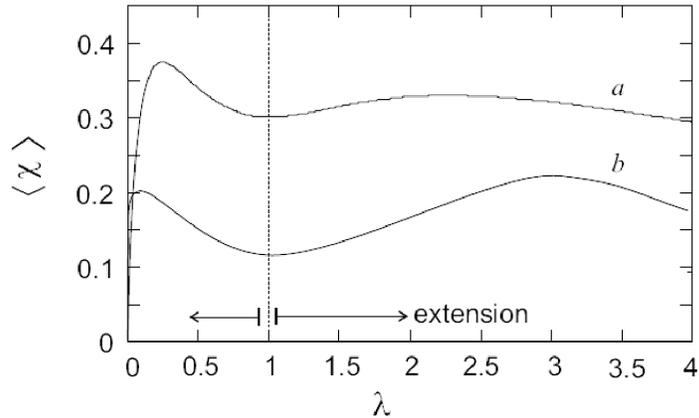}}
 \caption{The mean helical fraction of a network stretched by a
factor $\lambda$, according to ($8$) for $N = 100$ and
$\beta\Delta f = 0.05$ (curve a), $\beta\Delta f = 0.1$ (curve b,
for chains further away from the helix-coil transition).}
\label{meanhelical}
\end{figure}
This model of a single helix-forming polypeptide was later
extended to predict the stimulated helix-coil transition occurring
in a crosslinked random network of such chains under external
deformation \cite{Kutter2002}. Within a basic network theory,
assuming the affine deformation of random strands through their
endpoints and performing the quenched average over the network
topology, the average helical content (per chain) is given by
\begin{equation}
\langle{\chi{\rangle}}(\underline{\underline{\bm{\lambda}}}) =
\int  d\bm{R} \
\chi(|{\bm{{\underline{\underline{\lambda}}{R}}}}|) P({\bm{R}}),
\label{eq:chi}
\end{equation}
where $\underline{\underline{\bm{\lambda}}}$ is the affine
volume-conserving strain tensor and the quenched probability
distribution of finding a network strand with a given number of
monomers $N$ and end-to-end distance at formation $\bm{R}$ is
given by $P({{R}}) \propto \exp [-F_{ch}(R)/k_BT]$. In a random
network, depending on their orientation, some strands are extended
and some are compressed; the average helical content
$\langle{\chi{\rangle}}$, however, remains a non-monotonic
function of uniaxial extension as long as the chains are
``slightly denatured'' (above but close to their natural
helix-coil transition: $\Delta f$ positive but small).
Fig.~\ref{meanhelical} shows an example of this model predictions
for the specific case of uniaxial extension of the sample by a
factor $\lambda$. It can be seen that the average helical content
in the network increases on uniaxial extension (as well as on
compression, due to incompressibility) and goes through a maximum
at a certain $\lambda$.


\section{Experimental Methods}

We used gelatin networks as a model system. Gelatin gels are good
candidates for such experiments, as their constituting chains are
mostly in the denatured (coil) state and the samples can be
stretched with their optical activity simultaneously measured.
Moreover, gelatin gels are well described in the literature
\cite{Joly2002} and are by far the most studied functional
biopolymer due to its extensive practical use. The gelatin network
is held together by effective crosslinks made of right-handed
(tertiary) super-helices stabilized by hydrogen bonds, resulting
from the wrapping of three left-handed helical segments of
otherwise denatured collagen chains.

\subsection{Sample preparation}
\begin{figure}
\resizebox{0.6\textwidth}{!}{\includegraphics{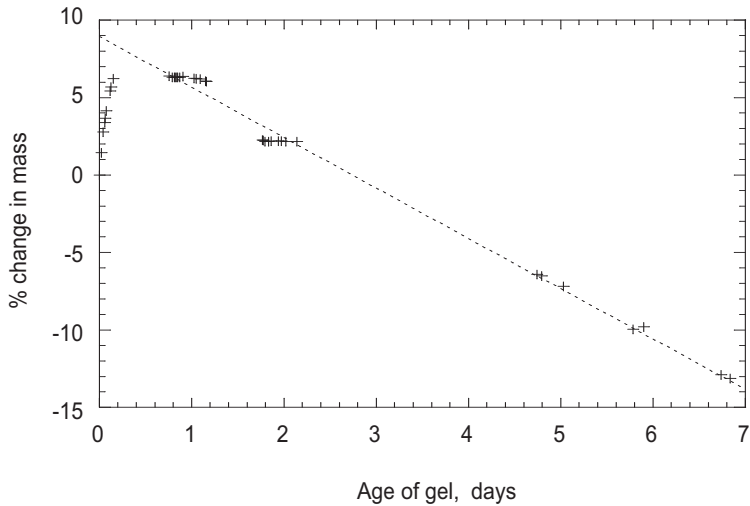}}
 \caption{Mass change an ethylene glycol gelatin gel at room
temperature in the air. The mass rises initially, then shows a
linear decrease at a rate of $\sim 3.3\%$ per day.} \label{weight}
\end{figure}

\begin{figure}
\resizebox{0.6\textwidth}{!}{\includegraphics{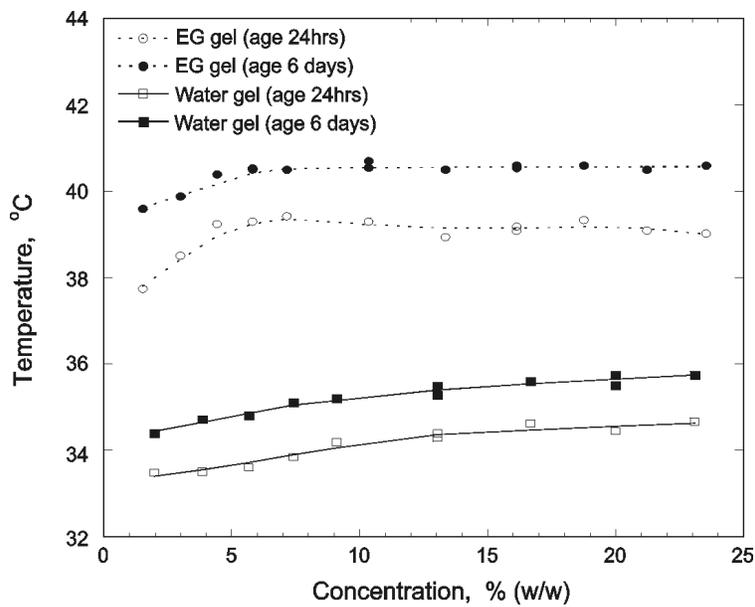}}
\caption{Melting point, T$_{m}$, as a function of concentration,
obtained from DSC, for ethylene glycol (EG) and water gelatin gels
aged at room temperature for 24 hours and six days.  Respective
curves are labelled on the plot.} \label{phase}
\end{figure}

\begin{figure}
\resizebox{0.5\textwidth}{!}{\includegraphics{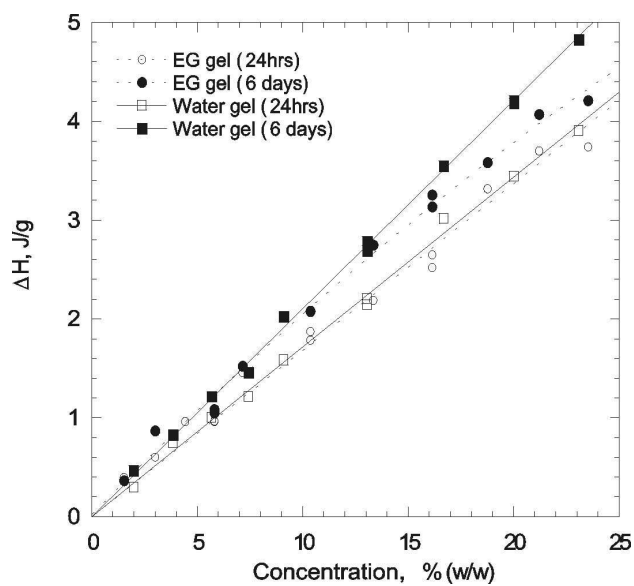}}
\caption{Transition enthalpy, $\Delta$H, as a function of
concentration, obtained from DSC for ethylene glycol (EG) and
water gels aged at room temperature for 24 hours and six days.
Respective curves are labeled on the plot.} \label{deltah}
\end{figure}

Gelatin samples were prepared by dissolving gelatin powder (from
Sigma) in ethylene glycol. The main reason for using ethylene
glycol instead of more traditional water is to allow us conduct
experiments over long periods of time and not worry about solvent
evaporation (boiling point $T_{B}\approx 190^{\circ}$). As we
indicate below, there are good reasons to believe that the
intrinsic chain responses are qualitatively the same in ethylene
glycol, although the specific magnitudes of elastic modulus and
helix content at any given temperature may be slightly different.
Figure~\ref{weight} shows the percentage change in the mass of
gelatin gel in ethylene glycol. As the gel consolidates, its mass
increases initially (presumably due to equilibrium hydration in
the atmosphere) and then shows a weak linear decrease. The rate of
this decrease is slow enough on the time scale of our experiments
(typically $30$ min) to have no significant effect on our results.
The sol-gel transitions of water and ethylene glycol (EG) gels
have been investigated using differential scanning calorimetry
(DSC), using a Perkin Elmer Pyris 1 calorimeter. The gels, kept in
air-tight aluminium pans, were heated at a rate of
5$^{\mathrm{o}}$C/min from 20 to 65$^{\mathrm{o}}$C. The melting
point, T$_{m}$, of the gel was determined from the peak of the DSC
endotherm. Figure~\ref{phase} shows T$_{m}$ as a function of
concentration for water and ethylene glycol gels, aged at room
temperature for one day and six days after the initial mixing. The
melting temperature increases as the gel ages and is approximately
5$^{\mathrm{o}}$C higher for ethylene glycol gels, compared to
water gels.
Figure~\ref{deltah} shows that the increase in the transition enthalpy
for the whole sample, $\Delta$H, with polymer concentration is almost
perfectly linear and is roughly the same for water and ethylene glycol
gels aged for one day for concentrations below 24\%(w/w), and for gels
six-days old below 10\%(w/w). Above this concentration the enthalpy
change is slightly greater for water gels.
Assuming the increased transition temperatures seen in ethylene
glycol gels are simply due to the greater viscosity of EG (with
finite rate of temperature change in the DSC), the density of the
crosslinks formed by the gelatin molecules in water and ethylene
glycol gels could be very similar. To determine the effect of
gelatin concentration, stretching experiments were performed using
$12\%$ and $16\%$(w/w) EG gels.

One important point about sample reproducibility has to be made.
Collagen from different sources, as well as samples prepared in
different batches, always differ in their properties. We did not
specifically undertake a task of making all samples uniformly
comparable. All reported effects are qualitatively reproduced,
however, small quantitative differences would always be observed.
Naturally, when samples had to be quantitatively compared (such as
in aging, or crosslinking density analysis), they were always
prepared from the same batch and at the same time.

\begin{figure}
\resizebox{0.6\textwidth}{!}{\includegraphics{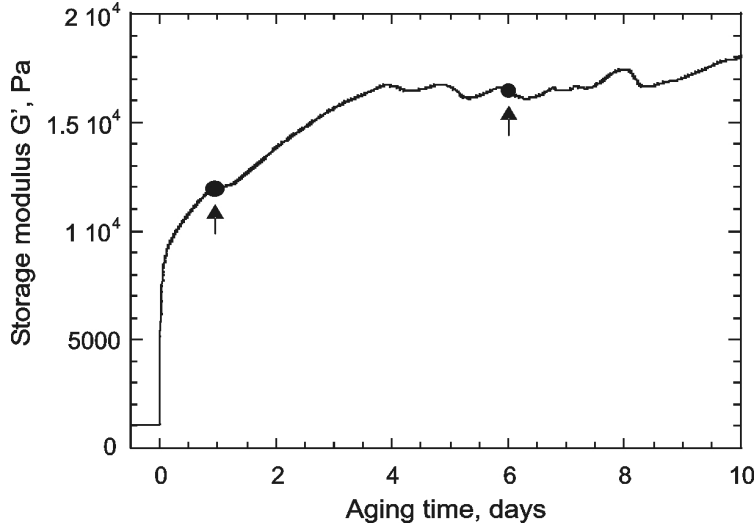}}
\caption{Variation of the storage modulus $G'$ with time after
quenching the 16\% EG gel to room temperature (at $t=0$ point).
$G'$ rises over four days to reach an equilibrium value of
$1.7\times 10^{4}$ Pa. Arrows point at the two aging states at
which main experiments have been performed (see text).}
\label{modulus}
\end{figure}

The mixture was kept above the gel transition point at
65$^{\mathrm{o}}$C under continuous stirring to ensure homogeneity
and consistent thermo-mechanical history of each experiment. After
cooling down to room temperature, each time a crosslinked gel
sheet of dimension $1.5\times{7}\times{20}$~mm was prepared. The
mechanical properties of the gelatin quenched below its gel point
have been investigated using a Dynamic Stress Rheometer
(Rheometrics Ltd). Figure~\ref{modulus} shows the variation in
linear storage modulus, $G'$, over $10$ days after quenching the
16\% EG gel to room temperature. The modulus rises initially when
the elastic network held together by the triple-helix crosslinks,
reaching a constant value of $1.7\times{10^4}$ Pa after four days.
At this point we assume the crosslinking is completed (leaving
aside delicate issues of slow drift of collagen towards its
natural state, irrelevant on our time scales). To determine the
effect of the variation in $G'$ on the optical rotation,
stretching experiments were performed one day after the gel was
removed from the mould, when the storage modulus was still rising,
and after six days, when $G'$ had reached equilibrium; both aging
points are marked on the plot.

\subsection{Opto-mechanical apparatus}
\begin{figure}
\resizebox{0.7\textwidth}{!}{\includegraphics{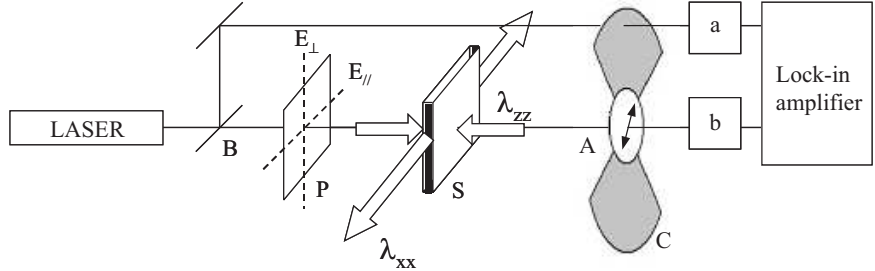}}
\caption{Scheme of the apparatus which combined optical rotation
rate $\partial\Psi/\partial z$ and stress-strain
$\sigma=f(\lambda)$ measurements. The components are labeled: (B)
beam splitter, (P) polarizer, (S) sample mounted, using clamps,
between a temperature compensated force transducer and micrometer
screw gauge controlled by a stepper motor [not shown], (A)
analyzer, (C) light chopper rotating on the same frame as (A).}
\label{apparatus}
\end{figure}

The opto-mechanical experiments have been carried out by combining a
dynamical method for measuring the optical rotation $\Psi$ and a
stress-strain apparatus, Fig.~\ref{apparatus}.  Gelatin networks,
typically of dimension $1.5\times{7}\times{20}$~mm, were mounted using
clamps between a temperature compensated force transducer (Pioden
Controls Ltd.~) and a micrometer screw gauge controlled by a stepper
motor.  The force transducer allowed for measurements of stress
$\sigma=f(\lambda)$ as the sample was strained along the
$\lambda_{xx}$ axis.  Samples were stretched at a strain rate of
$0.001$ s$^{-1}$.  Simultaneously, we measured the optical rotation
$\Psi$ of the gelatin gel with a linearly polarized light at $633$~nm
from a He-Ne laser. The reason for choosing this long wavelength is to
stay away from any molecular absorption band characteristic of
circular dichroism (CD).  That method of detecting protein (and other)
helices is totally inappropriate for our study because the
interpretation of CD signal is notoriously ambiguous.  Any distortions
of the helices, such as bending, can have a significant affect on the
CD spectrum.  It would certainly be inapplicable in our case when the
helices are distorted by the network deformations.  Our method of
choice is the optical rotation of plane-polarized light in
non-absorbing region, which can be demonstrated to be linearly
proportional to the concentration of helically correlated regions of
chains.

\begin{figure}
\resizebox{0.5\textwidth}{!}{\includegraphics{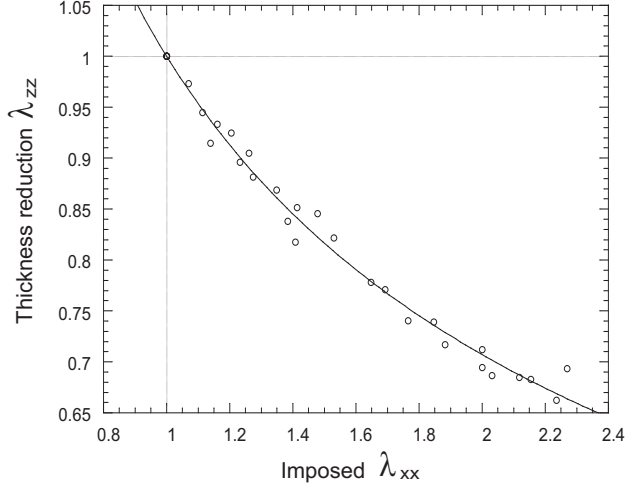}}
\caption{Variation of $\lambda_{zz}$ with $\lambda_{xx}$. The
curve corresponds to the transversely isotropic incompressible
$\lambda_{zz}=1/\sqrt{\lambda_{xx}}$.} \label{thickness}
\end{figure}

As shown in Fig.~\ref{apparatus}, the laser beam is split into two
parts, one plane-polarized beam (labeled b in
Fig.~\ref{apparatus}) passing through the sample and an analyzer,
rotating at a fixed frequency $\sim 16$~Hz, the other (beam a)
through an optical chopper mounted on the same rotor and providing
a reference signal. The lock-in amplifier (Stanford Research)
locked on the reference signal and detected the relative phase
difference $\Delta\Theta$ between the two signals. The phase shift
corresponds to the rotation of the plane of the linearly polarized
light and gives directly the total rotation angle $\Psi$. The
helix content is a material function proportional to the rate of
optical rotation $\partial\Psi/\partial z$, so we need to take
into account the change of path length of light $d$ (along $z$,
the sample thickness) as a function of the imposed uniaxial
extension $\lambda= \lambda_{xx}$. If we model the deformation as
strictly volume conserving, then
$\det(\mathbf{\underline{\underline{\lambda}}}) =
{\lambda_{xx}}{\lambda_{yy}}{\lambda_{zz}} = 1$.  If the
deformation is also isotropic, $\lambda_{yy} = \lambda_{zz}$ and
we expect $\lambda_{zz} = {1/\sqrt{\lambda_{xx}}}$. To verify
this, the width of a sample was measured using a traveling
microscope as a function of $\lambda_{xx}$. Figure~\ref{thickness}
shows that, on stretching the gel, $\lambda_{zz}$ and
$\lambda_{xx}$ accurately satisfy the predicted
$(1/\sqrt{\lambda})$ relation and hence the measured raw optical
rotation of a sample was corrected to give the rate of optical
rotation $\partial\Psi/\partial z \propto\Psi
\sqrt{\lambda}$. The measured stress was also corrected for the
corresponding decrease in the cross-sectional area (in the $yz$
plane) on stretching. Then the true stress, $\sigma'$, is related
to the measured nominal stress by $\sigma' =\lambda \,  \sigma$.

\subsection{Effect of induced birefringence}

The changes in rotation may be affected by birefringence, which is
necessarily induced in the network on uniaxial stretching.  The
gel is initially isotropic, but on extension the statistical
distribution of the polymer becomes biased. This is a very well
known effect observed in any polymer network under uniaxial
deformation. The induced birefringence is defined as $\Delta{n} =
n_{SD}-n_{ND}$ where $n_{SD}$ and $n_{ND}$ are the refractive
indices in the stretching and the normal directions. In our case
this effect was measured separately with the aid of a quarter-wave
plate inserted in the beam path of our apparatus, using a method
based on the phase differences of coherent light and therefore was
insensitive to intensity changes due to, for example, scattering.
Figure \ref{biref} shows the induced birefringence, $\Delta{n}$,
which is almost a linear function of imposed extension
$\lambda_{xx}$, for a 16\% gelatin sample.  It can be shown
\cite{Doi1986} that the induced birefringence is proportional to
the difference in the normal stress, $\Delta{n} =
C(\sigma_{zz}-\sigma_{xx})$, with $C$ the so-called stress optical
coefficient, which depends only on the local structure of the
polymer.  For our gelatin gel we determine $C \approx 1.6 \times
10^{-6}$cm$^2$/N, which agrees well with the literature data on a
variety of polymer networks, from dry rubber to highly swollen
gels \cite{Rennar1998,Hahn1997}.

One could in principle observe a rotation of the plane of
polarization, related to the induced birefringence $\Delta{n}$, if
the initial light polarization falls on the sample at an oblique
angle to the extension axis. The corresponding rotation angle
$\theta \sim \pi\Delta{n}\lambda_{zz}{d}/\Lambda$, where $\Lambda$
is the laser wavelength and $d$ the sample thickness. At our
highest extension and other parameters, we may obtain the value
$\theta_{max} \sim 6^{\circ}$. We shall see that this value is
almost an order of magnitude below our readings. However, there is
a second factor making the effect of induced birefringence
irrelevant. In our case we deliberately send the incident light
with polarization aligned with the axes of deformation. If the
sample was not optically active, there would be no rotation at
all. As the plane of polarization starts its rotation away from
this initial position, as the light travels along its path, the
effect of birefringence kick in. The birefringence simultaneously
induced in the sample acts to rotate the plane of polarization
back towards the axes of deformation. That would reduce the
amplitude of the perceived optical rotation, by the amount
significantly less than $\theta_{max}$.  As the main effect
described in this paper is a dramatic increase in the magnitude of
optical rotation on stretching, we conclude that the induced
birefringence has little effect.

\section{Results and Discussion}

\begin{figure}
\resizebox{0.6\textwidth}{!}{\includegraphics{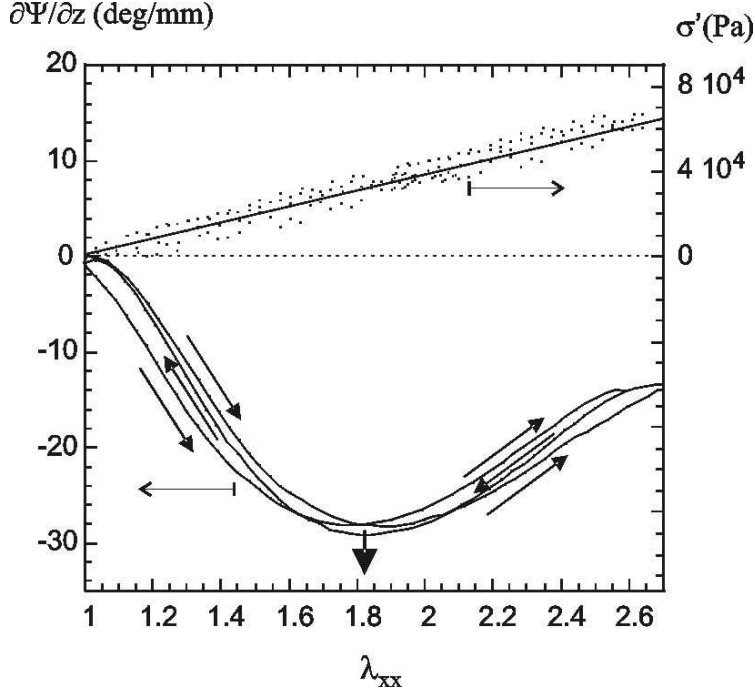}}
\caption{Evolution of the rotation rate $\partial\Psi/\partial z$
(left axis) and stress $\sigma '$ (right axis) as functions of
imposed $\lambda_{xx}$, for a 16\% gel.  The plot shows the data
of the extension, retraction and second extension cycles of the
sample, to demonstrate the reversible equilibrium nature of the
response.} \label{fig7a}
\end{figure}

\begin{figure}
\resizebox{0.5\textwidth}{!}{\includegraphics{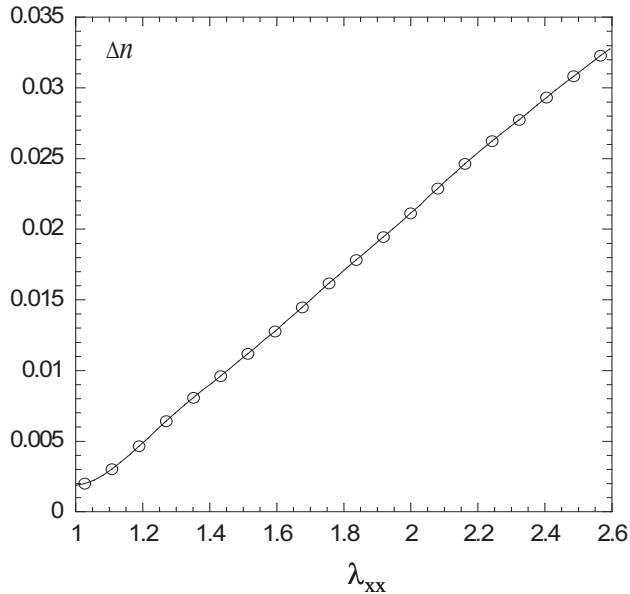}}
\caption{Birefringence, $\Delta{n}$, as a function of imposed
$\lambda_{xx}$, for a 16\% gel.} \label{biref}
\end{figure}

The equilibrium content of secondary helices in a crosslinked gel
changes as a function of the applied stress.  By combining
dynamic-mechanical and optical measurements, figure~\ref{fig7a} shows
the changes in optical rotation rate, $\partial\Psi/\partial
z$, which are a direct measure of the helical content, as a
function of imposed extension $\lambda_{xx}$. This result, generic
to this gelatin system (i.e. reproducing qualitatively over a wide
range of concentrations and temperatures), demonstrates that by
imposing an extension on chains, one can stimulate the helix
formation in the network strands.

The initial optical rotation of the unstretched network is
negative, $\partial\Psi/\partial z(\lambda=1)\approx
-0.63^{\circ}$/mm, which in our setup corresponds to left-handed
rotation. The optical rotation of a gelatin solution at this
concentration, at 50$^{\mathrm{o}}$C, when the chains are in the
random coil state, is an order of magnitude smaller ($\approx
-0.06^{\circ}$/mm) and still left-handed.  The optical rotation
response of native collagen has been compared to that of synthetic
peptides such as poly-proline. In aqueous solution, proline
polypeptides adopt the poly-L-proline II conformation, which is a
single left-handed helix, almost identical to the collagen
left-handed secondary helix \cite{Djabourov1988}. The optical
rotation of poly-L-proline II solutions shows similar behavior to
that observed for native collagen solutions \cite{blout}.  The
association of the helices to form the tertiary triple helix
marginally distorts the helical folded regions of chains, but has
little effect on the rotatory properties. The optical rotation
measurement is, therefore, mainly sensitive to the left-handed
secondary helical structure of individual chains.  The small
negative rotation measured initially (in the undistorted gelatin
sample, $\lambda=1$) is due to these secondary $\alpha$-helices,
which are joined into triple-helical regions that form the
crosslinks in the gel. Some additional single helices may be
present as well, but cannot be distinguished from those forming
the triple helices.

On extension the absolute value of the rotation rate increases
dramatically. At the same time, over the large range of extensions
$\lambda_{xx}$, the stress-strain relationship remains strictly
linear and fully reversible.  From this we may infer that the
total amount of network crosslinks (right-handed triple helix
regions) remains constant over the duration of this experiment and
the non-monotonic variation of optical rotation with extension is
not due to structural changes of the triple helical crosslinks.
The dramatic increase of the optical activity suggests an increase
of left-handed single-strand $\alpha$-helix population. These
helices are generated in the originally coiled network strands and
produce the increasing left-handed rotation (represented as
negative values of rotation rate in Fig.~\ref{fig7a}).

At higher extension, $\lambda_{xx}$, the variation of the optical
rotation is clearly non-monotonic and we found a maximum value of
its amplitude $-28.8^{\circ}$/mm for $\lambda^*\approx 1.85$, as
shown by the arrow in Fig.~\ref{fig7a}.  We assume this
corresponds to the maximum content of helices in the network,
which subsequently start to unwind for strain $\lambda>2$.  This
crossover is a consequence of the change in the nature of the
helix coil transition which becomes a helix-extended coil
transition for strong forces \cite{Varshney2005}.  The minimum in
the optical rotation could reflect the point at which the rotation
due to the induced birefringence dominates that resulting from the
formation of helices, leading to an increase in the optical
rotation. However, as the magnitude of the rotation due to
birefringence is small compared to the measured rotation, even at
our highest extension, it is likely that the result of the
birefringence is to shift the measured critical extension to a
somewhat smaller value than the true extension at which the
helices begin to unwind.

\begin{figure}
\resizebox{0.5\textwidth}{!}{\includegraphics{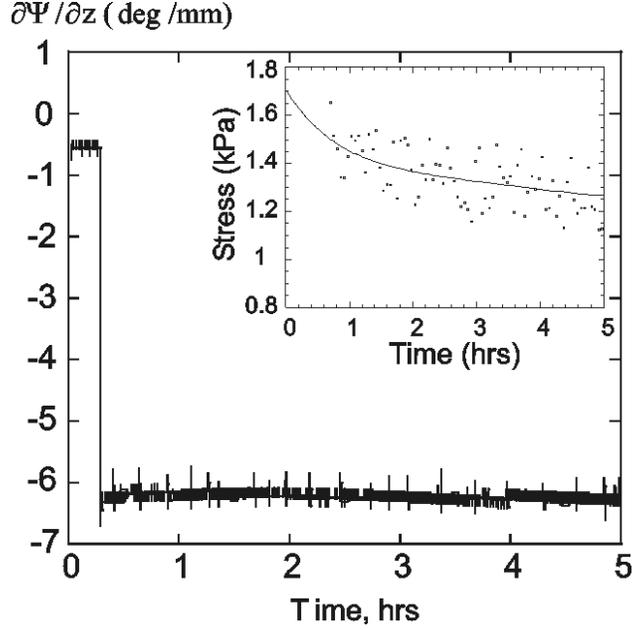}} \caption{Time
evolution of rotation rate after a strain step of $\sim 20\%$,
illustrating the stability of the microstructure. The inset shows
the corresponding slow stress relaxation, which is a well-known
feature of any rubbery network.} \label{fig7b}
\end{figure}

Our interpretation of the non-monotonic left-handed rotation depends
on the assumption that the conformation of the triple helical
crosslinks in the network is preserved; this is illustrated by
performing a repeated stretching-contracting cycle experiment. In
Fig.~\ref{fig7a} only an insignificant hysteresis is observed. This,
and the consistently linear stress-strain variation, suggests that the
change in average helical content on stretching is not due to
topological rearrangement of triple-helical junction zones (even
though a decrease of their density in the network could have been
initially suspected). To test the equilibrium in a different way, an
unstretched sample was extended very rapidly to $\lambda = 1.2$. The
optical rotation and stress were then measured over several hours
(Fig.~\ref{fig7b}). It is quite clear that the effects observed while
stretching the sample at a slow rate of $10^{-3}s^{-1}$ are close to
equilibrium.

We therefore conclude that the imposed strain promotes the induced
helix-coil transition on originally randomly coiled collagen segments
of the elastic network, which results in a large and non-monotonic
variation of the overall optical activity.

\subsection{Influence of crosslinks density}
\begin{figure}
\resizebox{0.85\textwidth}{!}{\includegraphics{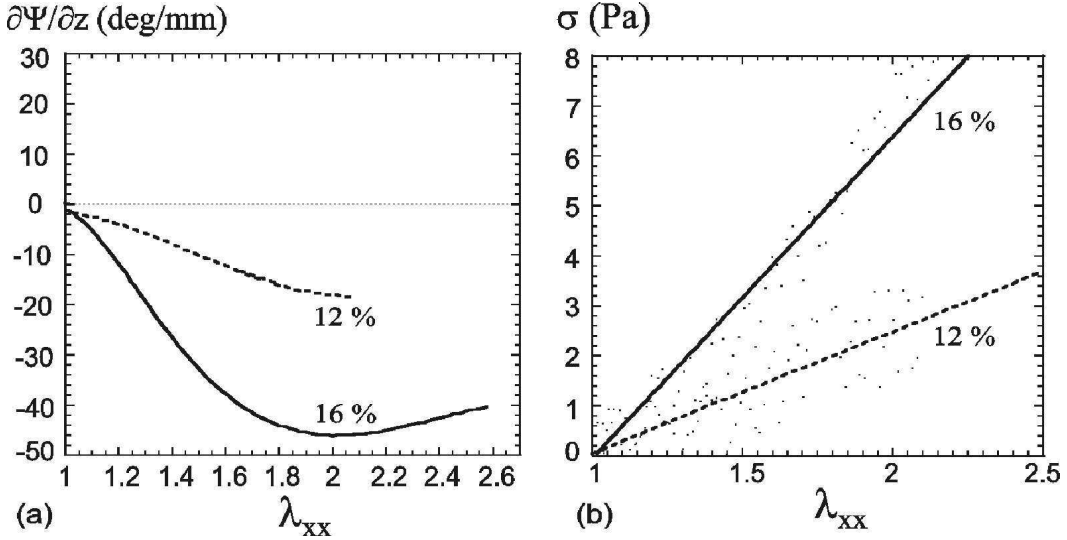}}
\caption{(a) Variation of the optical activity
$\partial\Psi/\partial z$ for $12\%$ and $16\%$ gels on
stretching, six days after the gels were removed from the mould. \
(b) The corresponding stress-strain measurements.} \label{fig8}
\end{figure}

 The influence of
the network properties on the stimulated helix-coil transition has
been studied by measuring the generic response described above as
a function of the crosslink density. The crosslink density at a
fixed temperature is directly related to the concentration of
gelatin and two different samples have been prepared with $12\%$
and $16\%$ of gelatin (w/w). Figure~\ref{fig8} shows the results
of stretching experiments performed on a $12\%$ (w/w) and a $16\%$
(w/w) gelatin gel (aged for six days after quenching, when the
storage modulus $G'$ had reached equilibrium, see
Fig.~\ref{modulus}). {The rates of optical rotation
$\partial\Psi/\partial z$ of unstretched (at $\lambda_{xx}=1$)
$12\%$ and $16\%$ samples are $-0.40^{\circ}/$mm and
$-0.63^{\circ}$/mm, respectively}. This is consistent with the
expectation that fewer triple helical crosslinks will form in a
$12\%$ gel. On stretching, the variations in optical rotation for
both $12\%$ and $16\%$ gelatin gels are non-monotonic with the
imposed deformation $\lambda_{xx}$. However, a difference is
observed on the magnitude of $\partial\Psi/\partial z$, while the
critical strain, $\lambda^{*}\approx 1.85$ is similar for both
samples. For a $12\%$ gel, the rotation rate reaches a maximum
value of approximatively $-18^{\circ}$/mm, smaller than the value
of $\partial\Psi/\partial z=-46.4^{\circ}$/mm for the $16\%$ gel.
On the corresponding stress-strain curves we find the consistently
linear increase with the extension $\lambda_{xx}$, but the value
of extension (Young) modulus $E$ for a $12\%$ gelatin gel is
approximately one-third of those for a $16\%$ gel, again
consistent with the expected reduction in the number of
crosslinks.

\begin{figure}
\resizebox{0.85\textwidth}{!}{\includegraphics{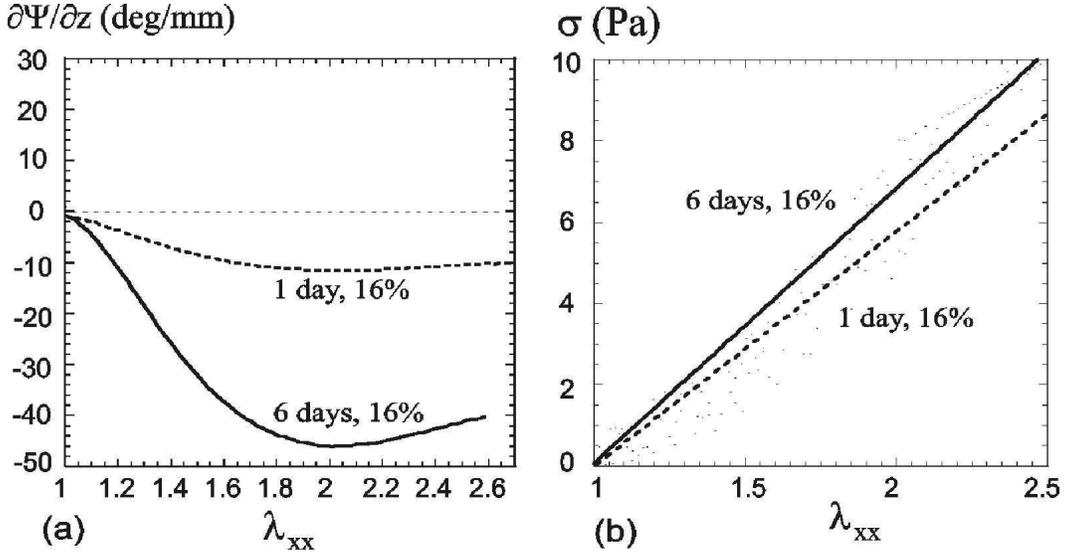}}
\caption{(a) Variation of the optical activity
$\partial\Psi/\partial z$ for a $1$-day and a $6$-day aged $16\%$
gel. \ (b) The corresponding stress-strain measurements showing
only minor variation in the modulus.} \label{fig9}
\end{figure}

\subsection{Influence of network aging}
 We examined how the
gelatin aging influences the mechanically stimulated helix
generation in the network. For this purpose, experiments have been
performed as a function of the storage modulus $G'$ evolution:
after $1$ day from gel quenching when $G'$ is still rising and
after $6$ days when $G'$ reaches a plateau (constant value of
$1.7\times 10^{4}$ Pa) from which we assume that the equilibrium
crosslinking is completed. The results are presented in
Fig.~\ref{fig9} which shows the variation of $\partial\Psi /
\partial z$ as a function of $\lambda_{xx}$ for a $16\%$ gelatin
gel after one day and six days of aging from quenching. For a
1-day old gel, the initial optical rotation for the unstretched
sample is $\partial\Psi/\partial z=-0.58^{\circ}$/mm for the
$16\%$ gel and $-0.38^{\circ}$/mm for the $12\%$ gel. This is
smaller but close to the optical activity of a 6-day old gel, at
each concentration, which is consistent with their similar
extension modulus values. In both cases, on the corresponding
stress-strain curves, we find a linear relationship as shown on
Fig.~\ref{fig9}(b). The values of the Young modulus $E$ (by
assuming the Poisson's ratio $r=0.5$), calculated from the
stress-strain relation in Fig.~\ref{fig9}, compare favorably with
those obtained using the dynamic rheometer: one expects, and
indeed finds $E \approx 3G'$. As the gel ages the helical content
increases and the network becomes stiffer.  This is a result of
the lengthening of the right-handed triple helices, reported and
discussed in, e.g., \cite{Colby2003}.

On stretching, the rate of optical rotation for an 1-day old gel
becomes negative and shows the same characteristic non-monotonic
variation. However, the value of the optical activity at the critical
strain $\lambda^{*}$ is much lower than the value obtained for a 6-day
old gel: for instance, for the $16\%$ gel the maximum values of
right-handed rotation rate are $\partial\Psi/\partial z=(-12.0\pm
0.5)^{\circ}$/mm and $\partial\Psi/\partial z=(-46.4\pm
0.5)^{\circ}$/mm, respectively.  This, with the results obtained as a
function of cross-links density, confirms that the induced helical
content depends strongly of the elastic properties of the network,
which has to transmit deformation to individual collagen strands in
order to induce their helices.

\subsection{Optical anisotropy of induced helices.}
\begin{figure}
\resizebox{0.5\textwidth}{!}{\includegraphics{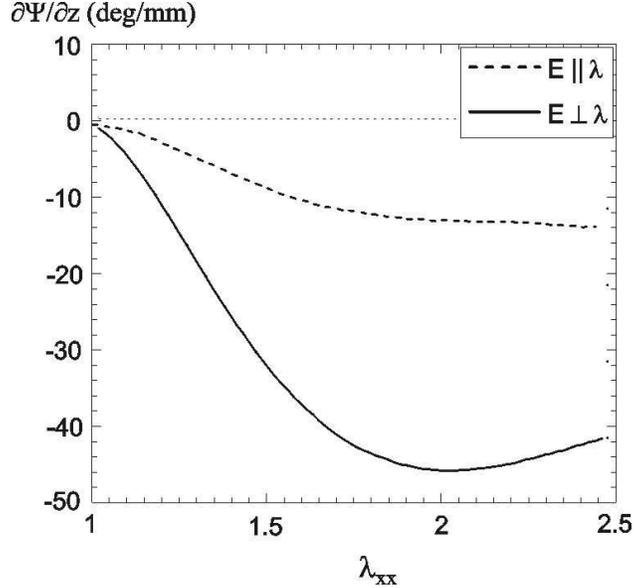}} \caption{The
optical activity $\partial\Psi/\partial z$ for a $6$-day aged
$16\%$ gel, sampled with the light,  linearly polarized parallel
and perpendicular to the axis of uniaxial stretching.}
\label{fig15n}
\end{figure}

 We have
further investigated the uniaxial anisotropy of induced helices
with sending the light through the stretched sample, which was
polarized perpendicular (electrical field $E\perp \lambda_{xx}$)
and parallel ($E\parallel \lambda_{xx}$) to the direction of
extension.  The results for a 6-day old 16\% gel are shown in
Fig.~\ref{fig15n}.

For the light initially polarized perpendicular to the direction
of extension, the rate of optical rotation is consistently greater
than that obtained when the polarization plane is initially
parallel to $\lambda_{xx}$.  The difference between the rotation
rates in these parallel and perpendicular geometries increases as
the strain is increased up to the critical strain. Qualitatively
similar results are obtained for 12\% gels and for 1-day old gels.
This effect is a signature of an optical anisotropy of induced
helices.

The peptide bond is optically anisotropic on itself
($\Delta\alpha_{0} = 2 \mathrm{x} 10^{-30}$ m$^{3}$)
\cite{Bras1998}.  When amino acids are arranged in a straight
$\alpha$-helix, the average uniaxial anisotropy along the helical
axis is clearly induced (in a different language, this is called
the transversely isotropic symmetry of the local dielectric
tensor). Assuming the helix is a relatively rigid structure, at
least in comparison with the coil chain regions, one expects their
alignment induced by the uniaxial strain applied to the system.
This alignment of an induced helix along the force axis is also a
feature of the original models of individual chains
\cite{Pincus2001,Buhot2002}. A simple affine model calculating the
average uniaxial anisotropy induced in the orientational
distribution of rigid rods embedded in an elastic continuum, which
is uniaxially stretched by a factor $\lambda$, predicts the
distribution function
 \begin{equation}
 P(\theta) = \frac{\lambda^3}{2(\cos^2\theta + \lambda^3
 \sin^2\theta)^{3/2}} \sin \theta \, d\theta ,
 \end{equation}
with $\theta$ the local angle of a helix axis to the direction of
strain. This has to be multiplied by the (yet unknown) factor of
anisotropy of helix polarizability, and then convoluted with the
non-monotonic function describing the helix concentration, cf.
Fig.~\ref{meanhelical} and section~\ref{sec:theo}, to account for
the observed effect of light polarization.

\section{Conclusions}
We performed a similar numerical integration to that of Kutter and
Terentjev \cite{Kutter2002} to obtain the effective optical
rotation rate, by weighting the (scalar) helical fraction by the
orientational bias of incident light polarization with respect to
the applied deformation. As would be expected, for a fixed helical
fraction, the rotation rate is greater for the light initially
polarized perpendicular to the helix axis, suggesting the negative
dielectric anisotropy. We applied this model to the gelatin
network used in this investigation.  The value of the parameter
$N$, the average number of chain segments between crosslinks, is
not straightforward to estimate, as the effective crosslinks do
not form at the ends of the molecule. The distance between
crosslinks $\xi$ could be estimated from the measured shear
modulus, $G \approx \frac{k_BT}{\xi{^3}}$, assuming basic rubber
elasticity response. For a $16\%$ six days old gelatin gel, this
gives $\xi\sim 6\times 10^{-9}$m. Since the dimensions of an
aminoacid are of the order of a few angstr$\ddot{\textrm{o}}$ms,
this suggests $N\sim 100$ for the gelatin network. In order to fit
the position of the rotation rate maximum, we required the
parameter $\beta\Delta f = 0.05$, at $N=100$, cf.
section~\ref{sec:theo}. Assuming the optical rotation linearly
increases with increasing helical content, this model shows
qualitatively the same variation with extension as the results
obtained with the gelatin gel.

In summary, the structural changes occurring in a biopolymer network
(gelatin gel) on extension have been monitored by observing changes in
its optical activity. We have demonstrated experimentally the
mechanical induced helix-coil transition by applying an external
deformation (imposed end-to-end distance) of the network.  The
initially small negative rotation of the plane of polarization by the
gel, due to the right-handed triple helical crosslinks, becomes large
on stretching and shows a distinct non-monotonic variation with the
imposed strain. This non-linear variation is not due to birefringence
and depends strongly on the elastic properties of the network. This
result, in qualitative agreement with theoretical predictions for
$\beta \Delta f =0.05$ and $N=100$ gives a consistent picture of
induced helix-coil transition occurring in polymer networks under
stress.  However, the dependence of the observed changes in optical
rotation on the plane of polarization remains to be explained.

\noindent {\small We thank EPSRC for financial support.
Discussions with R. Colby, A. Craig, S. Kutter, and the help of K.
Lim with many of the measurements, are gratefully appreciated.}



\begin{thebibliography}{15}
\expandafter\ifx\csname
natexlab\endcsname\relax\def\natexlab#1{#1}\fi
\expandafter\ifx\csname bibnamefont\endcsname\relax
  \def\bibnamefont#1{#1}\fi
\expandafter\ifx\csname bibfnamefont\endcsname\relax
  \def\bibfnamefont#1{#1}\fi
\expandafter\ifx\csname citenamefont\endcsname\relax
  \def\citenamefont#1{#1}\fi
\expandafter\ifx\csname url\endcsname\relax
  \def\url#1{\texttt{#1}}\fi
\expandafter\ifx\csname
urlprefix\endcsname\relax\def\urlprefix{URL }\fi
\providecommand{\bibinfo}[2]{#2}
\providecommand{\eprint}[2][]{\url{#2}}

\bibitem{cell}
\bibinfo{author}{\bibfnamefont{D.} \bibnamefont{Boal}},
\emph{\bibinfo{title}{Mechanics of the Cell}}
(\bibinfo{publisher}{Cambridge University Press, Cambridge},
\bibinfo{year}{2002}).

\bibitem[{\citenamefont{MacKintosh et al.}(1995)\citenamefont{MacKintosh et al}}]{MacKintosh1995}
\bibinfo{author}{\bibfnamefont{F.C.} \bibnamefont{MacKintosh}},
\bibinfo{author}{\bibfnamefont{J.} \bibnamefont{K\"{a}s}} \bibnamefont{and}
  \bibinfo{author}{\bibfnamefont{P.A} \bibnamefont{Janmey}},
  \bibinfo{journal}{Phys. Rev. Lett.} \textbf{\bibinfo{volume}{75}},
  \bibinfo{pages}{4425} (\bibinfo{year}{1995}).


\bibitem[{\citenamefont{Weitz}(1999)\citenamefont{Weitz et al}}]{Weitz1999}
\bibinfo{author}{\bibfnamefont{T.} \bibnamefont{Gisler}} \bibnamefont{and}
  \bibinfo{author}{\bibfnamefont{D.A} \bibnamefont{Weitz}},
  \bibinfo{journal}{Phys. Rev. Lett.} \textbf{\bibinfo{volume}{82}},
  \bibinfo{pages}{1606} (\bibinfo{year}{1999}).

\bibitem[{\citenamefont{Zimm et~al.}(1959)\citenamefont{Zimm and Bragg}}]{Zimm1959}
\bibinfo{author}{\bibfnamefont{B.H.} \bibnamefont{Zimm}} \bibnamefont{and}
  \bibinfo{author}{\bibfnamefont{J.K.} \bibnamefont{Bragg}},
  \bibinfo{journal}{J. Chem. Phys.} \textbf{\bibinfo{volume}{11}},
  \bibinfo{pages}{526} (\bibinfo{year}{1959}).

\bibitem[{\citenamefont{Grosberg et~al.}(1994)}]{Grosberg1994}
\bibinfo{author}{\bibfnamefont{A.Y.} \bibnamefont{Grosberg}}
\bibnamefont{and} \bibinfo{author}{\bibfnamefont{A.R.}
\bibnamefont{Khokhlov}}, \emph{\bibinfo{title}{Statistical Physics of Macromolecules}}
  (\bibinfo{publisher}{AIP Press, New York}, \bibinfo{year}{1994}).

\bibitem[{\citenamefont{Poland et~al.}(1970)}]{Poland1970}
\bibinfo{author}{\bibfnamefont{D.} \bibnamefont{Poland}}
\bibnamefont{and} \bibinfo{author}{\bibfnamefont{H.A.}
\bibnamefont{Scheragag}}, \emph{\bibinfo{title}{Theory of Helix-Coil Transitions in Biopolymers}}
  (\bibinfo{publisher}{Academic Press, New York}, \bibinfo{year}{1970}).

\bibitem[{\citenamefont{Perkins et~al.}(1994)\citenamefont{Perkins et al}}]{Perkins1994}
\bibinfo{author}{\bibfnamefont{T.T.} \bibnamefont{Perkins}},
\bibinfo{author}{\bibfnamefont{S.R.} \bibnamefont{Quake}}, \bibinfo{author}{\bibfnamefont{D.E.} \bibnamefont{Smith}} \bibnamefont{and}
  \bibinfo{author}{\bibfnamefont{S.} \bibnamefont{Chu}},
  \bibinfo{journal}{Science} \textbf{\bibinfo{volume}{264}},
  \bibinfo{pages}{822} (\bibinfo{year}{1994}).

\bibitem[{\citenamefont{Li et~al.}(1998)\citenamefont{Li et al}}]{Hongbin1998}
\bibinfo{author}{\bibfnamefont{H.} \bibnamefont{Li}},
\bibinfo{author}{\bibfnamefont{M.} \bibnamefont{Rief}}, \bibinfo{author}{\bibfnamefont{F.} \bibnamefont{Oesterhelt}} \bibnamefont{and}
  \bibinfo{author}{\bibfnamefont{H.E.} \bibnamefont{Gaub}},
  \bibinfo{journal}{Adv. Mater.} \textbf{\bibinfo{volume}{3}},
  \bibinfo{pages}{316} (\bibinfo{year}{1998}).

\bibitem[{\citenamefont{Rief et~al.}(1997)\citenamefont{Rief et
al}}]{Rief1997} \bibinfo{author}{\bibfnamefont{M.} \bibnamefont{Rief}},
\bibinfo{author}{\bibfnamefont{F.} \bibnamefont{Oesterhelt}},
\bibinfo{author}{\bibfnamefont{B.} \bibnamefont{Heymann}}
\bibnamefont{and} \bibinfo{author}{\bibfnamefont{H.E.}
\bibnamefont{Gaub}}, \bibinfo{journal}{Science}
\textbf{\bibinfo{volume}{275}}, \bibinfo{pages}{1295}
(\bibinfo{year}{1997}).

\bibitem[{\citenamefont{Li et~al.}(1998)\citenamefont{Li et al}}]{Oesterhelt1999}
\bibinfo{author}{\bibfnamefont{F.} \bibnamefont{Oesterhelt}},
  \bibinfo{author}{\bibfnamefont{M.} \bibnamefont{Rief}} \bibnamefont{and}
  \bibinfo{author}{\bibfnamefont{H.E.} \bibnamefont{Gaub}},
  \bibinfo{journal}{Adv. Mater.} \textbf{\bibinfo{volume}{3}},
  \bibinfo{pages}{316} (\bibinfo{year}{1998}).

\bibitem[{\citenamefont{Rief et~al.}(1997)\citenamefont{Rief
      etal}}]{Rief1998} \bibinfo{author}{\bibfnamefont{M.}
\bibnamefont{Rief}}, \bibinfo{author}{\bibfnamefont{J.M.}
\bibnamefont{Fernandez}}, \bibnamefont{and}
\bibinfo{author}{\bibfnamefont{H.E.}  \bibnamefont{Gaub}},
\bibinfo{journal}{Phys. Rev. Lett.}  \textbf{\bibinfo{volume}{81}},
\bibinfo{pages}{4764} (\bibinfo{year}{1998}).

\bibitem[{\citenamefont{Tamashiro et~al.}(2002)\citenamefont{Tamashiro and Pincus}}]{Pincus2001}
\bibinfo{author}{\bibfnamefont{M.N.} \bibnamefont{Tamashiro}} \bibnamefont{and}
  \bibinfo{author}{\bibfnamefont{P.} \bibnamefont{Pincus}},
  \bibinfo{journal}{Phys. Rev. E} \textbf{\bibinfo{volume}{63}},
  \bibinfo{pages}{021909} (\bibinfo{year}{2001}).

\bibitem[{\citenamefont{Buhot et~al.}(2002)\citenamefont{Buhot and Halperin}}]{Buhot2002}
\bibinfo{author}{\bibfnamefont{A.} \bibnamefont{Buhot}} \bibnamefont{and}
  \bibinfo{author}{\bibfnamefont{A.} \bibnamefont{Halperin}},
  \bibinfo{journal}{Macromolecules} \textbf{\bibinfo{volume}{35}},
  \bibinfo{pages}{3238} (\bibinfo{year}{2002}).


\bibitem[{\citenamefont{Kutter et~al.}(2002)\citenamefont{Kutter and Terentjev}}]{Kutter2002}
\bibinfo{author}{\bibfnamefont{S.} \bibnamefont{Kutter}} \bibnamefont{and}
  \bibinfo{author}{\bibfnamefont{E.M.}\bibnamefont{Terentjev}},
  \bibinfo{journal}{Eur. Phys. J. E} \textbf{\bibinfo{volume}{8(5)}},
  \bibinfo{pages}{539} (\bibinfo{year}{2002}).

\bibitem[{\citenamefont{Doi et~al.}(1986)}]{Doi1986}
\bibinfo{author}{\bibfnamefont{M.} \bibnamefont{Doi}}
\bibnamefont{and} \bibinfo{author}{\bibfnamefont{S.F.}
\bibnamefont{Edwards}}, \emph{\bibinfo{title}{The Theory of Polymer Dynamics}}
  (\bibinfo{publisher}{Oxford University Press, Oxford}, \bibinfo{year}{1986}).

\bibitem{Rennar1998}
\bibinfo{author}{\bibfnamefont{N.} \bibnamefont{Rennar}},
  \bibinfo{journal}{Phys. Chem. Chem. Phys.} \textbf{\bibinfo{volume}{102}},
  \bibinfo{pages}{1665} (\bibinfo{year}{1998}).

\bibitem{Hahn1997}
\bibinfo{author}{\bibfnamefont{O.} \bibnamefont{Hahn}},
\bibinfo{author}{\bibfnamefont{D.} \bibnamefont{Woermann}},
  \bibinfo{journal}{Phys. Cham. Chem. Phys.} \textbf{\bibinfo{volume}{101}},
  \bibinfo{pages}{703} (\bibinfo{year}{1997}).

\bibitem[{\citenamefont{Joly-Duhamel et~al.}(2002)\citenamefont{Joly-Duhamel}}]{Joly2002}
\bibinfo{author}{\bibfnamefont{C.} \bibnamefont{Joly-Duhamel}},
\bibinfo{author}{\bibfnamefont{D.} \bibnamefont{Hellio}} \bibnamefont{and}
  \bibinfo{author}{\bibfnamefont{M.} \bibnamefont{Djabourov}},
  \bibinfo{journal}{Langmuir} \textbf{\bibinfo{volume}{18}},
  \bibinfo{pages}{7208} (\bibinfo{year}{2002}).

\bibitem[{\citenamefont{Djabourov et~al.}(2002)\citenamefont{Djabourov}}]{Djabourov1988}
\bibinfo{author}{\bibfnamefont{M.} \bibnamefont{Djabourov}},
\bibinfo{author}{\bibfnamefont{J.} \bibnamefont{Leblond}} \bibnamefont{and}
  \bibinfo{author}{\bibfnamefont{P.} \bibnamefont{Papon}},
  \bibinfo{journal}{J. Phys. France} \textbf{\bibinfo{volume}{49}},
  \bibinfo{pages}{319} (\bibinfo{year}{1988}).

\bibitem{blout}
\bibinfo{author}{\bibfnamefont{E.R.} \bibnamefont{Blout}},
\bibinfo{author}{\bibfnamefont{J.P.} \bibnamefont{Carver}} \bibnamefont{and}
  \bibinfo{author}{\bibfnamefont{J.} \bibnamefont{Gross}},
  \bibinfo{journal}{J. Am. Chem. Soc.} \textbf{\bibinfo{volume}{85}},
  \bibinfo{pages}{644} (\bibinfo{year}{1963}).

\bibitem[{\citenamefont{Varshney et~al.}(2002)\citenamefont{Varshney}}]{Varshney2005}
\bibinfo{author}{\bibfnamefont{V.} \bibnamefont{Varshney}}
\bibnamefont{and} \bibinfo{author}{\bibfnamefont{G.A.} \bibnamefont{Carri}}
  \bibinfo{journal}{Macromolecules} \textbf{\bibinfo{volume}{38}},
  \bibinfo{pages}{780} (\bibinfo{year}{2005}).

\bibitem{Colby2003}
\bibinfo{author}{\bibfnamefont{L.} \bibnamefont{Guo}},
\bibinfo{author}{\bibfnamefont{R.H.} \bibnamefont{Colby}} \bibnamefont{and}
\bibinfo{author}{\bibfnamefont{A.M.} \bibnamefont{Howe}},
  \bibinfo{journal}{Macromolecules} \textbf{\bibinfo{volume}{36}},
  \bibinfo{pages}{10009} (\bibinfo{year}{2003}).

\bibitem [{\citenamefont{Bras et~al.}(1998)\citenamefont{Bras}}]{Bras1998}
\bibinfo{author}{\bibfnamefont{W.} \bibnamefont{Bras}},
\bibinfo{author}{\bibfnamefont{G.P.} \bibnamefont{Diakun}},
\bibinfo{author}{\bibfnamefont{J.F.} \bibnamefont{Diaz}},
\bibinfo{author}{\bibfnamefont{G.} \bibnamefont{Maret}},
\bibinfo{author}{\bibfnamefont{H.} \bibnamefont{Kramer}},
\bibinfo{author}{\bibfnamefont{J.} \bibnamefont{Bordas}},
\bibnamefont{and}
\bibinfo{author}{\bibfnamefont{F.J.} \bibnamefont{Medrano}}
\bibinfo{journal}{Biophys. J.} \textbf{\bibinfo{volume}{74}},
  \bibinfo{pages}{1509} (\bibinfo{year}{1998}).

\end{thebibliography}

\end{document}

\newpage
\begin{center} FIGURE LEGENDS \end{center}

FIG. 1: {} 

FIG. 2: {} 

FIG. 3: {} 

FIG. 4: {} 

FIG. 5: {} 

FIG. 6: {} 

FIG. 7: {} 

FIG. 8: {} 

FIG. 9: {} 

FIG. 10: {} 

FIG. 11: {} 

FIG. 12: {} 

FIG. 13: {} 

FIG. 14: { } 

\newpage

\end{document}